# REQUEST AND NOTIFICATION PATTERN FOR AN INTERNET BANKING SYSTEM

A. Meiappane[1], Dr. V. Prasanna Venkataesan[2]

[1] Research Scholar, [2] Associate Professor

[1,2] Pondicherry University, Puducherry, India – 605 014.

*Abstract:* **The quality of software is enhanced by using the design patterns. The design patterns are the reusable component used in the development of the software, which delivers improved quality software to the end users. The researchers have developed design patterns for user interface, e-commerce applications, mobile applications, text classification and so on. There are no design patterns for internet banking applications, but there is analysis pattern for banking. This motivated to mine the design patterns for internet banking application. It can be mined from the document of Business Process Management (BPM). In this paper the request and notification are two patterns, that have been presented, which have been mined from internet banking.**

## I. INTRODUCTION

The software architecture of a system is the set of structures needed to reason about them or structures of the system, which comprise software elements, the externally visible properties of those elements, and the relationships among them [1].

Architecture is the fundamental organization of a system embodied in its components, their relationships to each other and to the environment and the principles guiding its design and evolution.

### A. Software Architecture

In the field of computer science there occurs the complexity in choosing the data structure and the algorithm to implement the particular system. In order to rectify these complexities software architecture emerges. Design patterns, styles, description languages were developed during that time.

Software architecture reduces the complexity of abstraction and the separation of concerns. The software architecture deals with that of the concepts of components, connectors and styles. Software architecture seeks to build a bridge between business requirements and technical requirements by understanding use cases, and then finding ways to implement those use cases in the software. The goal of architecture is to identify the requirements that affect the structure of the application. Good architecture reduces the business risks associated with building a technical solution. A good design is sufficiently flexible to be able to handle the natural drift that will occur over time in hardware and software technology, as well as in user scenarios and requirements. An architect must consider the overall effect of design decisions, the inherent tradeoffs between quality attributes (such as performance and security), and the tradeoffs required to address user, system, and business requirements. The architecture should

- Expose the structure of the system but hide the implementation details.
- Realize all of the use cases and scenarios.
- Try to address the requirements of various stakeholders.

## II. PATTERN ORIENTED SOFTWARE ARCHITECTURE

When experts need to solve a problem, they invent a totally new solution. More often they will recall a similar problem they have solved previously and reuse the essential aspects of the old solution to solve the new problem. They tend to think in problem-solution pairs.

Identifying the essential aspects of specific problem - solution pairs leads to descriptions of problem - solving patterns that can be reused. The concept of a pattern as used in software architecture is borrowed from the field of (building) architecture. The goal of patterns within the software community is to help software developers resolve recurring problems encountered throughout all of software development. Patterns help create a shared language for communicating insight and experience about these problems and their solutions.

One of the widely used definitions is: "A pattern for software architecture describes a particular recurring design problem that arises in specific design contexts and presents a well proven generic scheme for its solution. The solution scheme is specified by describing its constituent components, their responsibilities and relationships, and the ways in which they collaborate".

*A. Patterns*
**Definitions:**
1. Giving solution to the recurring problem at the particular context[4].
2. Pattern falls into the family of similar problems. It is the process of distilling common factors from the system.
3. Pattern is said to be as the relation between the context, problem and solution.
4. A pattern system provides a pool of proven solutions to many recurring design problems. On another it shows how to combine individual patterns into heterogeneous structures and as such it can be used to facilitate a constructive development of software systems.
5. Patterns are the proven solution and not just a solution.
6. Patterns are useful, useable and used

Good patterns do the following:

- **It solves a problem:** Patterns capture solutions, not just abstract principles or strategies.
- **It is a proven concept:** Patterns capture solutions with a track record, not theories or speculation.
- **It describes a relationship:** Patterns don't just describe modules, but describe deeper system structures and mechanisms.
- **The pattern has a significant human component:** All software serves human comfort or quality of life; the best patterns explicitly appeal to aesthetics and utility.

**Qualities of Pattern**

1. **Encapsulation and Abstraction:** Each pattern encapsulates well defined problem and its solution in a problem domain.
2. **Openness and variability:** Each pattern should be open for extension or parameterization by other patterns so that they may work together to solve a larger problem.
3. **Generativity and Composability:** Each pattern once applied generates a resulting context which matches the initial context of one or more other patterns in a pattern language.
4. **Equilibrium:** Each pattern must realize some kind of balance among its forces and constraints.

*B. Elements of Pattern:*
Each pattern should contain the following elements [2]
**Pattern Name**
It is used to describe a design problem, its solutions and also its consequences. It increases design vocabulary. Pattern name is used to enhance the communication. Naming a pattern increases our design vocabulary. It should be a meaningful word or a single phrase.
**Context**
It is the situation giving rise to a problem. It is the description of the initial state before the patterns are applied. This should be a recurring situation.

**Problem**

A recurring problem arising in the situation. Describes when to apply the pattern. It explains the problem and its context. It might describe specific design problems such as how to represent algorithms as objects. It refers to the goal we are trying to achieve in this context and also refers to any constraint in the context. Sometimes the problem will include a list of conditions that must be met before it makes sense to apply the pattern

**Solution**

A proven resolution to the problem. Describes elements that make up the design, their relationships and their responsibilities. Does not describe specific concrete implementation. It tells about Abstract description of design problems and how the pattern solves it. It is a general design that anyone can apply which resolves the goal and set of constraints.

**Forces**

A description of the relevant forces and constraints and tells how they interact and conflict with one another and the goals we wish to achieve.

**Resulting Context**

The state of the system after the pattern has been applied. It tells about the consequences (both good and bad) of applying the pattern and also other problems and patterns that may arise from the new concept. This element describes which forces have been resolved and how and which remains unresolved.

**Examples**

One or more Sample applications of the pattern which tells a special initial context, how the pattern is applied to and transforms, that context, and the resulting context.

**Rationale**

A justifying explanation of steps or rules in the pattern and also of the pattern as a whole in terms of how and why it resolves its forces in a particular way to be in alignment with desired goals ,principles . Rationale provides insight into its internal workings.

**Related Patterns**

It is the static and dynamic relationships between this pattern and others within this same pattern language or system. Related patterns often share some common forces.

**Known Uses**

Describes known occurrences of the pattern and its application within the existing system.

*C. Types of Pattern*

- Architectural patterns
- Design Patterns
- Idioms

**Architectural Pattern**

It is said to be as the high level of patterns. Architectural pattern tells about the fundamental structural organization of the system. It provides the set of pre defined subsystems and specify their responsibilities and also includes rules and guidelines. In general Architecture pattern tells how to build the system.

**Design Patterns**

It is said to be as the medium level of patterns. Design Pattern is a general repeatable solution to a commonly occurring problem in software design. It refines the subsystem or the components. It gives the template of how to solve the problem. The design we obtain from the design pattern is directly transformed to the code [2].

**Idioms**

It is the low level of patterns. It describes how to implement the particular aspect of components. It addresses both the design and the implementation. It eases communication among the developers and speed up software development and maintenance. A collection of such related idioms defines the programming style.

**Motivation of Design Patterns**

- Designing reusable code is difficult. Patterns provide reusable frameworks and successful reusable design exists**.**
- It tells about the detailed design of the framework.

## III. BUSINESS PROCESS MODELING (BPM)

BPM is the process of representing processes of an enterprise, so that the current process may be analyzed and improved. BPM is typically performed by business analysts and managers who are seeking to improve process efficiency and quality. In this process we are going to improve the internet banking application so that we can able to get a reusable application from which lot of users feel ease to use the internet banking application[5].

### A. Pattern Identification

Architecture and design patterns, as demonstrated solutions to recurring problems, have proved practically important and useful in the process of software development. They have been extensively applied in industry. Discovering the instances of architecture and design patterns from the source code of software systems can assist the understanding of the systems and the process of re-engineering. More importantly it also helps to trace back to the original architecture and design decisions, which are typically missing for legacy systems [8].

Software architecture is the key artifact in software design, describing the main elements of a software system and their interrelationships. We present a method for automatically analyzing the quality of architecture by searching for architectural and design patterns from it.

In addition to approximating the quality of the design, the extracted patterns can also be used for predicting the quality of the actual system.

In this paper we are going to identify patterns in the internet banking application where the quality of the architecture is improved [6]. We are going to analyze the overall process in the internet banking and apply any pattern identification approach and we identify various patterns so that we can able to achieve reusability in our system.

### B. Pattern Identification Approach

In order to identify patterns some of the methods are available .Pattern identification is done by using the methods like similarity scoring , using multilayered approach(DeMIMA), through random walks[10], Among all these approaches the efficient method is the crosscutting concern method to identify patterns [9]

## IV. CROSSCUTTING CONCERN

Understanding of concerns plays an important role in successful software development. Modularization of concerns is important for software development. Object oriented programming paradigm provides an ease of modularization of basic concerns. There are some concerns whose implementation cannot be Modularized using object oriented paradigm like profiling, logging etc. The implementation of such concerns remains scattered throughout the source code. Such concerns are called crosscutting concerns [8]. Identification of crosscutting concerns plays an important role in aspect mining, defect detection and software maintenance.

Cross-cutting concerns are aspects of a program which affect other concerns and results in either scattering or tangling[11].

Every concern might not get modularized into a separate module. Such concerns whose implementation is scattered over more than one module are called crosscutting concerns. Such concerns lead to tangling of code.

Poor modularization of crosscutting concerns results in source code which has more defects and is difficult to maintain. For development of good quality software, it is essential to identify crosscutting concerns

Identification of crosscutting concerns at different stages of software development is essential for 3 reasons.
1. Refactoring legacy system to aspect oriented system.
2. For modularized implementation of the concern.
3. For appropriate distribution of testing effort error prone crosscutting code need to be identified.

In the internet banking application analysis on various processes[3] should be done on the internet banking in order to find the crosscutting approaches. Considering all these features the pattern should be identified.

## V. PATTERN MINING

Pattern mining is that mining of various patterns in the internet banking system by using BPM. It is used to represent the process involved in the internet banking and is used to analyze the process and then we apply crosscutting approach in this model.

The Crosscutting Concerns are the aspect of program which affects the other concerns of the program. These concerns cannot be decomposed from the rest of the system in both the design and the implementation and can result in either scattering , tangling .This crosscutting concern is used to improve the modularity of the system. The internet banking system as lot of services available to its end users. Hence we need to use the crosscutting concern approach in order to find different patterns available in the internet banking system.

There are various services involved in the internet banking like account, Third party transfer credit card, debit card, mutual fund, loan, insurance etc[7]. Based on the analysis a model is drawn for business process modeling in the internet banking.

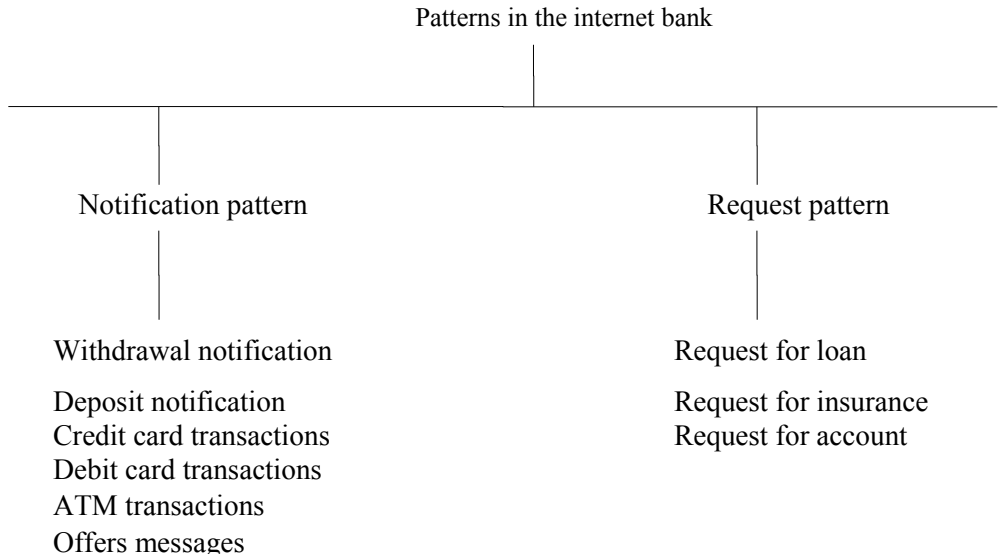

**Fig.1 : Patterns Mined from the Internet Banking System**

- **Notification pattern mined from Observer Pattern**

The notifications sent to the customers is specified in the patterns. The code is scattered in all types like Credit card, Debit Card, ATM. In all these situations we use the same coding with slight modifications. This account pattern is similar to that of the observer Pattern.

- **Request pattern Mined From the Chain Of responsibility pattern**

In the internet banking several request passing is done in cases like loan sanctioning, insurance sanctioning and in case of opening an account. In all these cases request is passed to different handlers until the request is fulfilled. In all these purposes coding is same. This request pattern is similar to that of the chain of responsibility pattern.

### A. Notification (Observer) Pattern

**Context**: The notification has to be done whenever there is a need. If any changes occurs in the system or if anything which needs to be intimated to the customers or stakeholders for this type of activities we do notifications.

**Problem:**
The modification or updating has to be intimated to the respective users who intend for it. This modification should be updated to the various modules and the classes.

**Solution:**
Here we use the observer pattern in order to notify the customers. Observer pattern is that if the observers are registered with the subject then the customers are notified about the changes. The notification is done in cased like informing telephone bill, loan interest due date to their customers who have registered.

Here we took the internet banking application where if any new services are added then the observers like (email ,SMS) will intimate about the changes to all the customers.

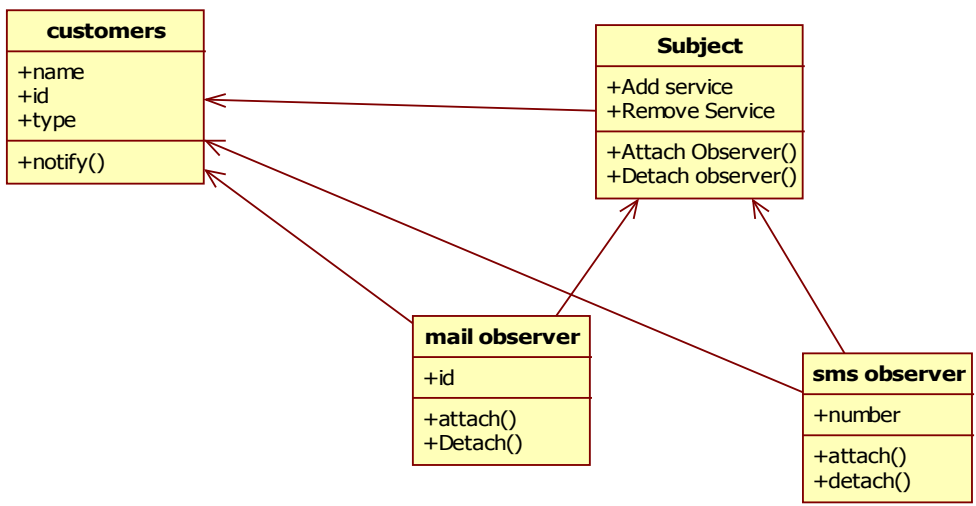

**Fig 2: Notification pattern class Diagram**

### B. Request(Chain Of Responsibility) Pattern

**Context:** Whenever we need to pass the request we use pattern. It tells about how we pass the request how the request is handled by the series of handlers and also how the request is completed without any interruption.

**Problem:**
This type of making the request is used in many type of the applications. We need to handle series of request by different handlers without any interruption.

**Solution:**
The request pattern is based on that of the chain of responsibility pattern. Here the process is passed through the different handlers until the request is completed without any of the interruption. The passing of the request is done in telephone repairing, passport, ration card requests etc

In the Internet Banking application The Requests for loan approving is processed as follows

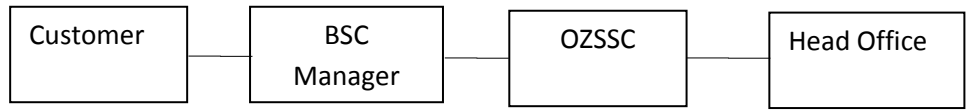

Customer passes the loan application to the BSC (Branch Section manager ) If he couldn't able to approve loan then the request is passed to OZSSC ( Operational Zonal Screening And Sanction Committee ) if he couldn't then the request is passed to HO (Head Office ). Hence the request is passed through different handlers.

The process of passing insurance application is as follows

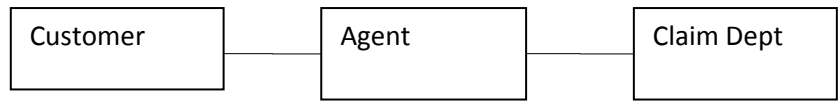

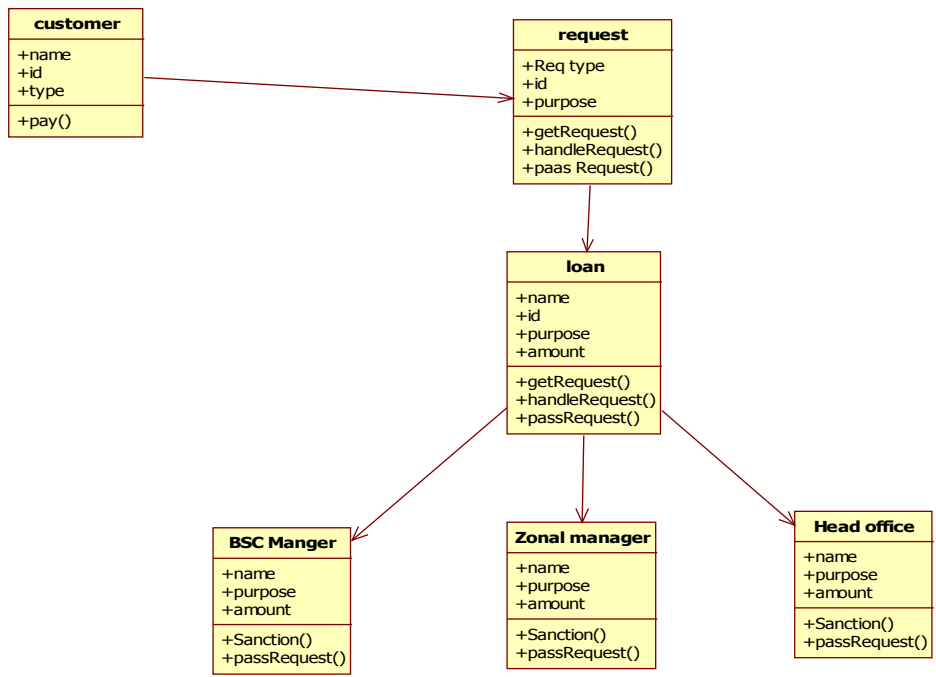

**Fig. 3 : Request pattern Class Diagram**

Any type of the request can be made of this type. Even in the case of insurance we can use this type of the process where we claim a certain amount and hence the request is passed through various handlers and finally the insurance amount will be sanctioned or the meeting is announced by the series of handlers.

The request pattern can be used in any kind of application in the case where the request needs to be passed through the series of handlers. Hence the request pattern has been discussed as above.

## VI. CONCLUSION & FUTURE ENHANCEMENT

Various patterns like Proxy Pattern, and Mediator pattern, have been identified. By implementing and using all these patterns we can able to achieve reusability in the internet banking application. It is more effective from the side of customers as well as web designer. By implementing this modularity is improved Thus the patterns are identified and achieve reusability in the internet banking application. After implementing the above mentioned patterns in the internet banking architecture we need to evaluate the system. We are proposing the SAAM evaluation so that we can compare the results of the pattern based system and the non pattern based system.

The future Enhancement is to evaluate the different architecture of the internet banking system by considering various quality attributes. By using the SAAM method t he existing system needs to be evaluated and also the pattern based system is evaluated . By assuming weights in both the cases the results are compared and we need to provide a efficient architecture of the System. The different architecture needs to be evaluated and the best one is chosen based on the evaluation results.